\documentstyle[aps,pra,epsf]{revtex}
\newcommand{\ket}[1]{\left | \, #1 \right \rangle}
\newcommand{\bra}[1]{\left \langle \, #1 \right | }
\title{Coherent manipulation of two dipole-dipole interacting ions}
\author{A.~Beige, S.F.~Huelga \cite{Huelga}, 
P.L.~Knight, M.B.~Plenio, and R.C.~Thompson}
\address{Optics Section, Blackett Laboratory, Imperial College London, 
London SW7 2BZ, England}
\date{\today}
\begin{document}

\maketitle

\begin{abstract}
We investigate to what extent two trapped ions
can be manipulated coherently when their coupling is mediated by a 
dipole-dipole interaction. We will show how the resulting level shift induced by this interaction can be used to 
create entanglement, while the decay of the states remains nearly negligible. 
This will allow us to implement conditional dynamics (a CNOT gate)
and single qubit operations. We propose two different experimental 
realisations where a large level shift can be achieved and discuss both the strengths and weaknesses of this scheme from the point
of view of a practical realization.
\end{abstract}

\vspace*{0.2cm}

\section{Introduction}
The practical implementation of quantum communication and quantum computation protocols \cite{steane,plerev} is a challenging task for experimental physics. Initial steps have already been accomplished. Conditional quantum dynamics
have been demonstrated using ultra cold trapped ions \cite{monroe,cz} and cavity QED schemes \cite{qed}, while NMR techniques have allowed the implementation of quantum algorithms involving a few 
qubits \cite{nmr}. However, the experimental demonstration of large scale protocols will, 
most likely, require a different technology.
In general, a successful candidate for implementing quantum logic must exhibit strong, coherent 
interaction between qubits and between each qubit and an external driving field \cite{PleKni2}. At the same time,
any other coupling to the external environment should be as weak as possible, in order to make the effect
of decoherence and dissipation negligible. 

A possible physical realization of quantum logic consists of making use of 
qubits whose coupling is mediated via dipole-dipole
interaction, as in the proposal by A.~Barenco et al.~\cite{adriano}
and the more recent proposals by G.K.~Brennen et al.~\cite{caves} and
D.~Jaksch et al.~\cite{cirac} using a dilute gas of neutral atoms. 
The aim of the present paper is to investigate how efficient the dipole-dipole interaction can be for quantum computing
using an experimental setup which is realizable with presently available technology. We will analyse to what extent two interacting two-level trapped ions can be used to perform conditional logic operations. 
Each single qubit is stored in an internal
electronic state, the ground state $|0 \rangle_i$ $(i=1,2)$ corresponds to the logic value 0 and the excited state $|1 \rangle_i$ to the value 1. 
We assume the distance of 
the ions $r$ to be small compared with the wavelength $\lambda_0$ of the 0-1 transition, 
e.g.~$2 \pi r = 0.2 \, \lambda_0$ or even smaller. To achieve this regime with conventional oscillation trap frequencies 
($\le 60\,$MHz) \cite{brian}, the wavelength $\lambda_0$ should be of the order of a few $\mu$m. This fact makes it very difficult to use a conventional ground state, but states with higher quantum numbers $n$ can provide an appropriate two-level system. The limitations of this configuration as far as the efficiency of the gate operation is concerned will be analysed in detail.

From the point of view of current experimental realisation, the necessity of cooling the ions to the motional ground state is one of the most limiting factors of the scheme proposed by Cirac and Zoller \cite{cirac}. In the scheme proposed here it is unnecessary to cool the ions to the motional ground state.
Therefore no assumption which excludes heating during the gate operations has to be made. The present proposal allows us to prepare entangled states of the two ions by applying a single laser pulse either in a running or in a standing wave configuration. We
will also show how to implement a controlled-not (CNOT) and single bit operations. 

In the following it will be convenient for us to use the basis ${\cal B}=\left\{ \ket{g}, \ket{e}, \ket{s},\ket{a} \right\} $, where
\begin{eqnarray} \label{11}
& & |g \rangle \equiv |00 \rangle,~~|e \rangle \equiv |11 \rangle, \nonumber \\
& & |s \rangle \equiv (|01 \rangle + |10 \rangle)/\sqrt{2} \nonumber \\
{\rm and} ~~ & & |a \rangle \equiv (|01 \rangle - |10 \rangle)/\sqrt{2}.
\end{eqnarray}
In this basis, the dipole-dipole interaction has two effects on the system. On the one hand, one has the well known fact that the ions can now decay faster or slower than two independent ions \cite{Dicke,Brewer2,Brewer1}, the decay rates being always smaller than $2A$, where $A$ is the Einstein coefficient of a single ion. On the other hand, the levels $\ket{s}$ and $\ket{a}$ are shifted with respect to their unperturbed energies as shown in Figure 1. This level shift is dependent on the distance between the ions \cite{MiKni}. The central idea is to choose the distance between the ions such that the resulting level shift is large enough to address all possible transitions shown in the scheme of Figure 1 separately.  This will allow us to create controlled entanglement between the ions. The use of the level shift to reduce stepwise excitations in an ion pair in favour of cooperative ``two-photon'' transitions was discussed some time ago \cite{Kim}.

\hspace*{1cm} \begin{figure}[htb]
\unitlength 1.2cm
\begin{picture}(13,5)
\thicklines
\put(4.55,2.5) {\line(1,0){1.5}}
\put(5.75,5) {\line(1,0){1.5}}
\put(5.75,1) {\line(1,0){1.5}}
\put(6.85,3.5) {\line(1,0){1.5}}
\put(3.2,3) {\line(1,0){0.5}}
\put(3.2,1) {\line(1,0){0.5}}
\put(9.1,3.5) {\line(1,0){0.5}}
\put(9.1,2.5) {\line(1,0){0.5}}
\thinlines
\put(4.55,3) {\line(1,0){0.3}}
\put(5.15,3) {\line(1,0){0.3}}
\put(5.65,3) {\line(1,0){0.3}}
\put(6.25,3) {\line(1,0){0.3}}
\put(6.85,3) {\line(1,0){0.3}}
\put(7.45,3) {\line(1,0){0.3}}
\put(8.05,3) {\line(1,0){0.3}}
\put(3.45,1) {\vector(0,1){2}}
\put(3.45,3) {\vector(0,-1){2}}
\put(9.35,2.5) {\vector(0,1){1}}
\put(9.35,3.5) {\vector(0,-1){1}}
\put(5.25,2.5) {\line(2,5){1}}
\put(6.75,1) {\line(2,5){1}}
\put(7.75,3.5) {\line(-2,3){1}}
\put(6.25,1) {\line(-2,3){1}}
\put (3.6,2){$\hbar\omega_0$}
\put (9.5,3){$\hbar {\rm Im}\,C$}
\put (7.3,0.95){$|g\rangle$}
\put (7.3,4.95){$|e\rangle$}
\put (8.4,3.45){$|s\rangle$}
\put (4.15,2.45){$|a\rangle$}
\end{picture}
\caption{A four level system describing two dipole-dipole interacting two level ions with energy separation $\hbar \omega_0$. The interaction leads to a shift of level $\ket{s}$ and level $\ket{a}$ which is proportional to the imaginary part of the separation dependent coupling constant $C$. Laser fields can be used to excite transitions between the four basis states $ \ket{g}$, $|s \rangle$, $|a \rangle$ and $\ket{e}$, as indicated by the thin lines.} \label{bild2}
\end{figure}
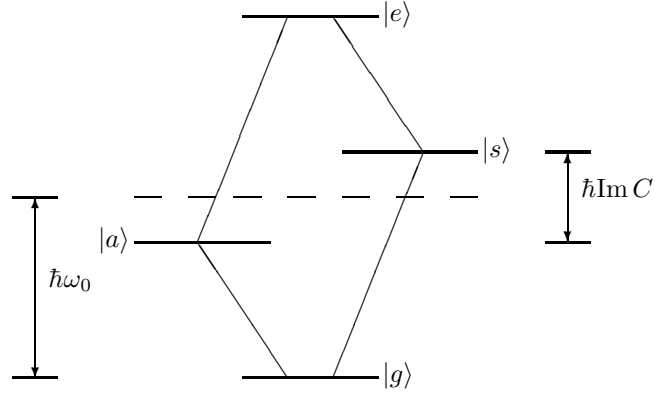

The paper is organised as follows. In Section II a short description of two dipole-dipole interacting ions is presented using the quantum jump approach \cite{HeWi,HeSo,PleKni,MC,QT,CohDal}. This method has been applied recently to the case of two ions in Refs.~\cite{BeHe4,BeHe5}. The necessary results will be summarised and the interesting parameter range for quantum computing is 
described. In Section III-V we present a simple algorithm to prepare the maximally entangled states of the two-qubit system, $|s \rangle$ and  $|a \rangle$, and show how a controlled NOT operation (CNOT) and single qubit operations can be performed.

\section{Description of the physical system.}

We consider two ions fixed at positions ${\bf r}_i$ and each with two levels, $|0\rangle_i$ and 
$|1\rangle_i$ with energy difference $\hbar \omega_0$.  Both ions are assumed to have the same 
dipole moment ${\bf D}_{01} = ~_i\langle 0|{\bf X}|1\rangle_i$ forming an angle $\vartheta$ with 
the line connecting the ions. The time evolution of the system is governed by the Hamiltonian 
\begin{eqnarray} \label{21}
H &=& \sum_{i=1,2} \hbar \omega_0 \, |1\rangle_{ii}\langle 1|
+ \sum_{{\bf k}\lambda} \hbar \omega_{\rm k} \, a^\dagger_{{\bf k}\lambda} a_{{\bf k}\lambda} + {\rm i}\hbar \sum_{i=1,2} \sum_{{\bf k}\lambda} 
\left( g_{{\bf k}\lambda} \, a_{{\bf k}\lambda} {\rm e}^{{\rm i} {\bf k} \cdot 
{\bf r}_i} \, \sigma_i^+ - {\rm h.c.} \right)
\end{eqnarray}
where $\sigma^\pm_i$ are the lowering and the raising operators 
acting on ion $i$, $|0 \rangle_{ii}\langle 1|$ and $|1 \rangle_{ii} \langle 0|$, respectively. 
The operators $a_{{\bf k}\lambda}$ and $a^\dagger_{{\bf k}\lambda}$ denote the free radiation field 
annihilation and creation operators of a photon in the mode $({\bf k},\lambda)$. The last term includes the coupling constant $g_{{\bf k}\lambda}$,
\begin{eqnarray}  \label{22}
g_{{\bf k}\lambda} &=& e \left( {\omega_k \over 2 \epsilon_0 \hbar L^3} \right)^{1/2} \left( {\bf D}_{01},{\bf \epsilon}_{{\bf k}\lambda} \right),
\end{eqnarray}
and describes the coupling strength between each ion and the quantised radiation field. $L^3$ corresponds to the quantisation volume. Going over to the interaction picture with respect to the free field and the atomic Hamiltonian, the interaction Hamiltonian becomes
\begin{eqnarray} \label{23}
H_{\rm I} = {\rm i}\hbar \sum_{i=1}^2 \sum_{{\bf k}\lambda} \left( \,
g_{{\bf k}\lambda} \, a_{{\bf k}\lambda} \,  
{\rm e}^{{\rm i} (\omega_0-\omega_k) t} \, 
{\rm e}^{{\rm i} {\bf k} \cdot {\bf r}_i} \, \sigma_i^+ 
- {\rm h.c.} \, \right).
\end{eqnarray}
The operator $H_{\rm I}$ already contains the dipole-dipole interaction of the two ions as seen 
from the master equations \cite{aga}, from the quantum jump approach \cite{BeHe4,BeHe5} or from much earlier work \cite{MiKni}. In the formulation used here this interaction is due to an exchange of virtual photons between the ions \cite{MiKni,aga}. The Coulomb interaction between the electric charges of the ions is not included in the Hamiltonian because this interaction is compensated by the trap potential in order to keep the ions at a fixed distance.

\subsection{The conditional no-photon time evolution for two two level atoms with dipole-dipole
interaction.}

The quantum jump approach allows us to derive from the interaction Hamiltonian in Eq.~(\ref{23}) 
the dynamics of a single two-ion system \cite{HeWi,HeSo,PleKni}. 
We now briefly summarise the main results. The time evolution of the system 
under the condition that no photon is emitted is described by the conditional non-Hermitian Hamiltonian $H_{\rm cond}$, which is obtained from the relation
\begin{eqnarray}  \label{24}
I\!\!I -{{\rm i} \over \hbar}\, H_{\rm cond} \, \Delta t 
&\cong  & \langle 0_{\rm ph} |U_{\rm I}(\Delta t,0)| 0_{\rm ph} \rangle
\end{eqnarray}
where $\Delta t$ has to satisfy $1/A \ge \Delta t \ge 1/\omega_0$.
The right hand side of Eq. (\ref{24}) can be evaluated in second order perturbation theory. 
For the two two-level ions one obtains \cite{BeHe4}
\begin{eqnarray} \label{25}
H_{\rm cond} &=& \frac{\hbar}{2{\rm i}} \left[ 
\, A \left( \sigma_{1}^+\sigma_{1}^- + \sigma_{2}^+\sigma_{2}^- \right) 
+ C \left( \sigma_{1}^+\sigma_{2}^- + \sigma_{2}^+\sigma_{1}^- \right) \,
\right]  
\end{eqnarray}
with the $r$ dependent coupling constant for electric dipole transitions
\begin{eqnarray} \label{26}
C &=& {3A \over 2} \, {\rm e}^{{\rm i} k_0 r} \Bigg[ \,
{1 \over {\rm i} k_0 r} \left( 1 - \cos^2 \vartheta \right) 
+ \left( {1 \over (k_0 r)^2} -{1 \over {\rm i}(k_0 r)^3} \right) 
\left( 1 - 3 \cos^2 \vartheta \right) \, \Bigg]
\end{eqnarray}
which describes the dipole-dipole interaction between the ions and where
$k_0 = 2\pi r/\lambda_0$. Here $\vartheta$ is defined through 
$\cos^2 \vartheta = \left| \left( {\bf D}_{01} , {\bf r} \right) \right|^2/r^2 D_{01}^2$ with ${\bf r} = {\bf r}_2 - {\bf r}_1$.
The dependence of $C$ on $r$ is maximal for $\vartheta = \pi/2$ and we will assume this value in our analysis in the following. 

Using the basis ${\cal B}$ of Eq. (\ref{11}) the Hamiltonian becomes
\begin{eqnarray} \label{27}
H_{\rm cond} &=& {\hbar \over 2 {\rm i}} \, \Big[ \,
(A + C) \, |s \rangle \langle s| + (A - C) \,
|a \rangle \langle a| + 2 A \, |e \rangle \langle e| \, \Big] 
\end{eqnarray}
which is diagonal.
As shown in \cite{BeHe4,BeHe5,aga}, $A \pm {\rm Re} \, C$ and $2A$ are the decay rates of the levels $|s \rangle$, $|a \rangle$, and $|e \rangle$, 
respectively. From Eq. (\ref{26}) one can check that ${\rm Re} \, C \le A$. Therefore the decay rates are always smaller than $2A$. The imaginary part of $C$ is related to the level shift of the states $|s \rangle$ and $|a \rangle$, respectively, which is due to the dipole-dipole interaction between the ions. For small distances ${\rm Im}\,C$ becomes much larger than the decay rate $A$ because, according to Eq.~(\ref{26}) and using the value $\vartheta=\pi/2$, one has
\begin{eqnarray} \label{28}
{\rm Im}\,C &=& -3A \left[ {\sin k_0r \over (k_0 r)^2} 
+ {\cos k_0r \over (k_0 r)^3} \right]~. 
\end{eqnarray}
The dependence of the imaginary part of the coupling constant $C$ on the distance between the ions $r$ is shown in Figure 2. 

\begin{figure}
\begin{center}
\input{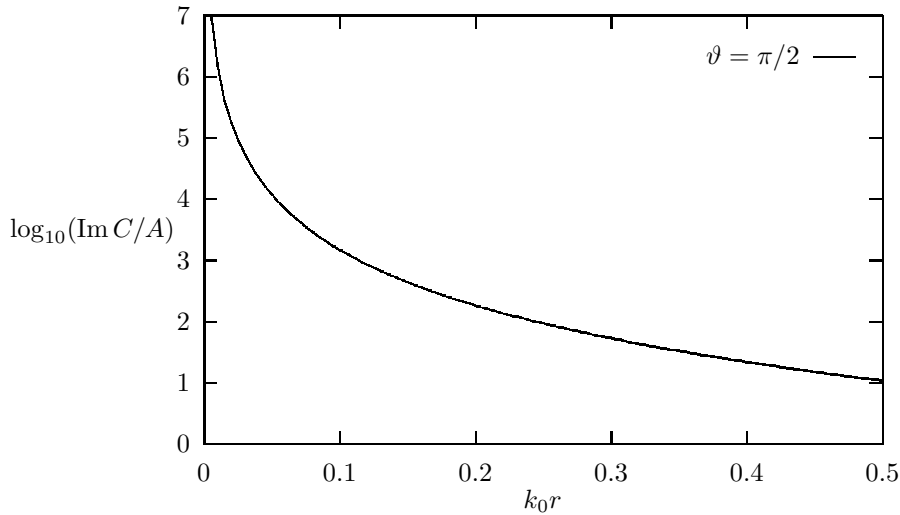}
\end{center}
\caption{Dependence of the cooperative shift ${\rm Im} \, C$ on $r$. For $k_0r<0.2$ the imaginary part of $C$ becomes much larger than $A$. Note that ${\rm Im} \, C$ is given in units of the Einstein coefficient $A$.}
\end{figure} 

\subsection{Suitable parameter range for coherent manipulation.}

The crucial parameter for the coherent manipulation of two dipole interacting ions is the cooperative level shift Im$\,C$. It will be used in the following to create entanglement between the ions and to perform  conditional logical operations. 
In this section we will discuss how to achieve a level shift large enough to allow the dipole-dipole interaction between ions to be the basic mechanism for creating entanglement. Let us suppose that $k_0 r=0.2$, r being the
ion spacing. This
yields a level shift Im$\,C=375\,A$ (See Figure 2). For a wavelength $\lambda_0=10$ $\mu $m, the resulting separation
between the ions is $r=0.318$ $\mu $m. Considering small deviations from equilibrium under the Coulomb force, the ion center-of-mass mode oscillation frequency can be estimated to be
\begin{equation} \label{trap}
{\omega \over 2 \pi} = {1 \over 2 \pi}
\sqrt{\frac{e^2}{2\pi \epsilon_0 M m_u \, r^3}} 
\approx 46 \, {\rm MHz}
\label{fre}
\end{equation}
for a system with atomic number $M=100$. In Eq.~(\ref{trap}) $m_u$ corresponds to the atomic mass unit. However, we have not been able to find dipole transitions of the required wavelength and involving the atomic ground
state. To achieve the required line shift forces us to use higher excited levels, as illustrated in Figure 3, where we explore the use of a Rydberg atom as an extreme case (this may be hard to realize in practice due to their sensitivity to electric stray fields). The qubit is stored in two Rydberg levels with $n \sim 20$ and the resulting gate operation time is limited to be much smaller than the characteristic decay rates
of the system (typically of the order of $1\,$ms). 

\begin{minipage}{6.54truein}
\begin{figure}[htb]
\begin{center}
\epsfxsize6.0cm
\centerline{\epsfbox{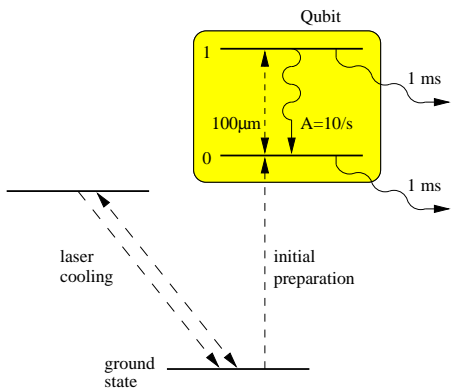}}
\end{center}
\caption{\small A large level shift can be achieved by using two excited levels with a transition wavelength of more than $10$ $\mu$m. The initial preparation should then involve pumping from the ground state to a higher quantum number $n$.} 
\end{figure}
\end{minipage}\\[.2cm]

A different scheme that we have considered consists of making use of shorter wavelengths. Figure 4 illustrates the relevant energy levels for ${\rm Yb}^+$. The qubit is now stored in two levels and the time scale for coherent manipulation is of the order of a few ms. As can be seen from Eq. (\ref{fre}), the required oscillation frequency would have to be at least $127\,$MHz $(k_0r=0.25$ and $\lambda_0=3.43\,\mu$m) or
$178\,$MHz $(k_0r=0.2)$, which exceeds the values that can be achieved in a conventional trap at the moment \cite{brian}. Nevertheless, no fundamental reason exists which precludes a large line shift, as described in the two previous schemes, although it would certainly be experimentally demanding.

\begin{minipage}{6.54truein}
\begin{figure}[htb]
\begin{center}
\epsfxsize10.0cm
\centerline{\epsfbox{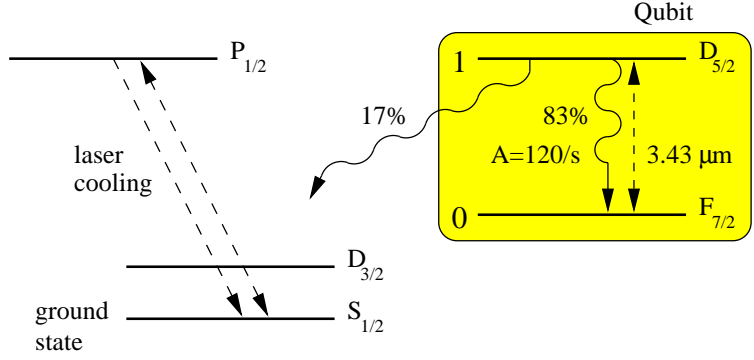}}
\end{center}
\caption{\small A different scheme for achieving a large level shift consists of using two levels with
a separation of the order of a few $\mu$m. We have depicted here the relevant energy levels for ${\rm Yb}^+$. Experimentally
this requires very large oscillation frequencies.} 
\end{figure}
\end{minipage}\\[0.2cm]

In a time scale $t \ll 1/A$, dissipation effects are negligible and the Hamiltonian $H_{\rm cond}$, which gives the time 
evolution of the ions, becomes in the Schr\"odinger picture
\begin{eqnarray} \label{29}
H_{\rm cond} &=& \hbar \omega_0 \Big( \, |s \rangle \langle s| 
+ |a \rangle \langle a| + 2 \, |e \rangle \langle e| \, \Big)
+ {\hbar \over 2} \, {\rm Im} \, C \Big( \,
|s \rangle \langle s| - |a \rangle \langle a| \, \Big) \equiv H_0.
\end{eqnarray}
Given that we neglected the decay of the ions, the Hamiltonian $H_{\rm cond}$ becomes Hermitian and we can omit the index ${\rm cond}$ from now on.
In the following we will analyse the dynamics of the two ions system within the interaction picture
where $H_0$ is the free Hamiltonian.
This has the advantage that in the absence of driving laser fields,
the state of the ions remains unchanged.

\section{Generation of maximally entangled states.}

In this Section we will describe how the maximally entangled states $|s \rangle$ and $|a \rangle$, one of the requisites for quantum computing, can be 
prepared just by applying a single laser pulse in a running wave or in a standing wave configuration,
respectively. We assume both ions to be initially in the ground state. 

Let us first consider a single laser field with frequency $\omega_{\rm L}$. The corresponding Hamiltonian describing the interaction with the laser, in the Schr\"odinger picture and in the usual rotating wave approximation, is given by
\begin{eqnarray} \label{31}
H_{\rm L} &=& e \sum_{i=1}^2 \Big( \, {\bf D}_{01} \cdot {\bf E}_0 \,
\sigma_i^- \, {\rm e}^{-{\rm i}({\bf k}_{\rm L} {\bf r}_i -\omega_{\rm L}t)} 
+ {\rm h.c.} \, \Big) \equiv \frac{\hbar}{2} \sum_{i=1}^2 
\Big( \Omega_i \, \sigma_i^- \, {\rm e}^{{\rm i} \omega_{\rm L} t} + {\rm h.c.} \Big)
\end{eqnarray}
with the Rabi frequencies 
\begin{eqnarray} \label{32}
\Omega_i &=& \frac{e}{\hbar} {\bf D}_{01} \cdot {\bf E}_0 
\, {\rm e}^{-{\rm i} {\bf k}_{\rm L} \cdot {\bf r}_i}.
\end{eqnarray}
Each ion sees a different Rabi frequency as a result of the small phase shift of the 
laser field due to the different positions of the ions. Here ${\bf E}_0$ and ${\bf k}_{\rm L}$ denote the 
electric field amplitude and the wave vector of the laser field, respectively. Going to 
the interaction picture with respect to the Hamiltonian $H_0$ of Eq. (\ref{29}),
the Hamiltonian of the whole system using the basis ${\cal B}$ of Eq. (\ref{11}) becomes
\begin{eqnarray} \label{34}
H_{\rm I} &=& \frac{\hbar}{2\sqrt{2}} \,
\Big[ (\Omega_1+\Omega_2) \Big( 
{\rm e}^{-{\rm i} {\rm Im}\,C t/2} \, |g \rangle \langle s| 
+ {\rm e}^{{\rm i}{\rm Im}\,C t/2} \, |s \rangle \langle e| \Big) \nonumber \\
& & - (\Omega_1-\Omega_2) \Big( 
{\rm e}^{{\rm i} {\rm Im}\,C t/2} \, |g \rangle \langle a| 
- {\rm e}^{-{\rm i} {\rm Im}\,C t/2} \, |a \rangle \langle e| \Big) 
\Big] \, {\rm e}^{{\rm i}(\omega_{\rm L}-\omega_0)t} + {\rm h.c.}  
\end{eqnarray} 
as one has
\begin{eqnarray}
\sigma_1^- &=& \ket{0}_{1\,1} \! \bra{1}  \otimes I\!\!I_2 
~=~ \Big( |g \rangle \langle s| - |g \rangle \langle a|
+ |s \rangle \langle e| + |a \rangle \langle e| \Big)/\sqrt{2}, \nonumber \\
\sigma_2^- &=& I\!\!I_1 \otimes \ket{0}_{2\,2} \! \bra{1} ~=~ \Big( |g \rangle \langle s| + |g \rangle \langle a|
+ |s \rangle \langle e| - |a \rangle \langle e| \Big)/\sqrt{2}.
\end{eqnarray} 
As can be seen from Eq. (\ref{34}), the laser can excite all four possible transitions in the two-ion 
system shown in Figure 1 separately. For instance, if both Rabi frequencies 
$\Omega_1$ and $\Omega_2$ are the same, only transitions between states of the same symmetry 
with respect to the exchange of the ions are excited. 
In addition, the energy of the states $|s \rangle$ and $|a \rangle$ is shifted with respect to their unperturbed values. These two features will be used in the following to generate entangled states and to perform conditional quantum logic.

\subsection{Generation of the symmetric state $|s \rangle$.}

To selectively excite the transition to level $s$, the driving laser should be in phase for both ions, e.g. $\Omega_1=\Omega_2$. This can be achieved by choosing ${\bf k}_{\rm L}$ orthogonal to the line joining the ions, and one obtains
\begin{eqnarray} \label{o1}
\Omega_1 &=& \frac{e}{\hbar} {\bf D}_{01} \cdot {\bf E}_0,
\end{eqnarray}
which we assume to be real, and where ${\bf E}_0$ is the amplitude of the laser at ion 1. In addition, the laser detuning $\omega_{\rm L} - \omega_0$ has to be equal to ${\rm Im} \, C/2$. Then the Hamiltonian $H_{\rm I}$ of Eq.~(\ref{34}) becomes
\begin{eqnarray} \label{36}
H_{\rm I} &=& \frac{\hbar\Omega_1}{\sqrt{2}} \, \Big( |g \rangle \langle s| 
+ {\rm e}^{{\rm i} {\rm Im}\,C t} \, |s \rangle \langle e| \Big) + {\rm h.c.}
\end{eqnarray} 
If the detuning of the $s$-$e$ transition is sufficiently large, e.g.
${\rm Im}\,C \ge 4 \, \Omega_1$ as in Figure 3, the second term in $H_{\rm I}$ gives only a minor contribution. Therefore, in analogy with the case
of the excitation of a single two-level system with a resonant laser field,
the population $P_{\rm s}$ of the state $|s \rangle$ at time $t$ is, to a good approximation, given by
\begin{eqnarray} \label{38}
P_{\rm s}(t) &=& \sin^2 \left(\Omega_1 t/\sqrt{2} \right).
\end{eqnarray} 
As a result, the maximally entangled state $|s \rangle$ can be prepared by simply applying a laser pulse with the duration 
\begin{eqnarray}
T_\pi &=& \pi/(\sqrt{2}\Omega_1),
\end{eqnarray}
i.e.~a standard $\pi$ pulse. 

The effect of the laser pulse taking into account the full Hamiltonian (\ref{36}) is shown in Figure 3, where the population of levels $g$, $s$ and $e$ has been determined numerically. Note that level $\ket{a}$ is not affected by the laser. Simulations show that the laser pulse prepares  to a very good approximation the entangled state $|s \rangle$ if the Rabi frequency is small compared with ${\rm Im} \, C$. The closeness of the prepared state to the ideal one is measured by the fidelity $F$. Here $F$ is equal to the obtained maximum population of level $s$. For $\Omega_1=0.25 \, {\rm Im}\, C$ one has $F=96\%$. For higher Rabi frequencies the term proportional ${\rm e}^{{\rm i}{\rm Im}\,Ct}$ cannot be considered as a negligible fast oscillating term and $\Omega_1$ has an upper bound. Therefore, the speed of
this operation cannot be made arbitrarily high. When $\Omega_1=0.25 \, {\rm Im}\,C$, it is $T_\pi=0.024/A\ll 1/A$ if one has $k_0r=0.2$. For the configuration of Figure 3 $T_\pi$ is only much smaller than the corresponding decay rate of the system if $k_0r \le 0.1$.

\begin{figure}
\begin{center}
\input{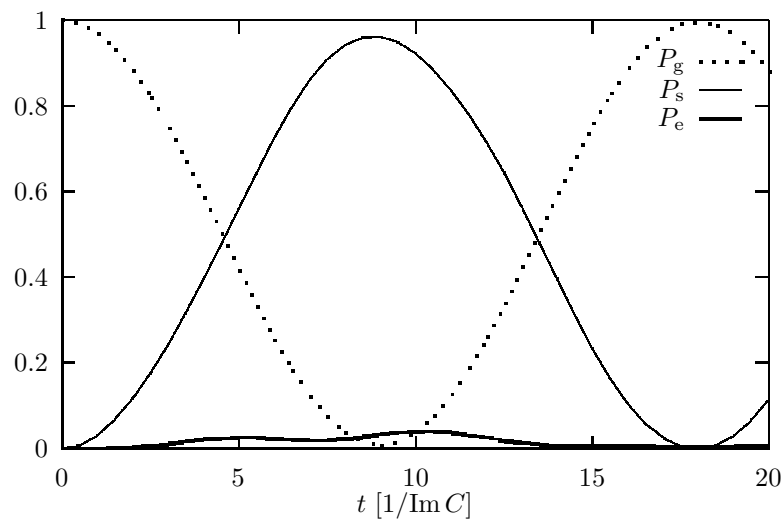}
\end{center}
\caption{Time dependence of the population of the levels $s$, $e$ and $g$ when
a single laser in phase with both atoms and in resonance with the $g$-$s$ transition is applied. 
The Rabi frequency is $\Omega_1=\Omega_2=0.25 \, {\rm Im}\,C$. 
The entangled state $|s \rangle$ can be prepared by a single laser pulse of duration $T_\pi$, while state $|a \rangle$ remains unaffected.}
\end{figure}

In a real experiment, the Rabi frequency of the driving laser may vary during the pulse duration or may simply not be known perfectly. We now assume that it is indeed equal to some value $\Omega$ instead of $\Omega_1$. In Figure 6 we have analysed the quality of the state preparation. This shows the population of state $|s \rangle$ after a time $T_\pi= \pi/(\sqrt{2}\Omega_1)$ with $\Omega_1=0.25 \, {\rm Im} \, C$ as a function of the ratio $\Omega/\Omega_1$. The fidelity $F$ is still high even for deviations of $\Omega$ from the assumed value $\Omega_1$ close to $20 \%$.

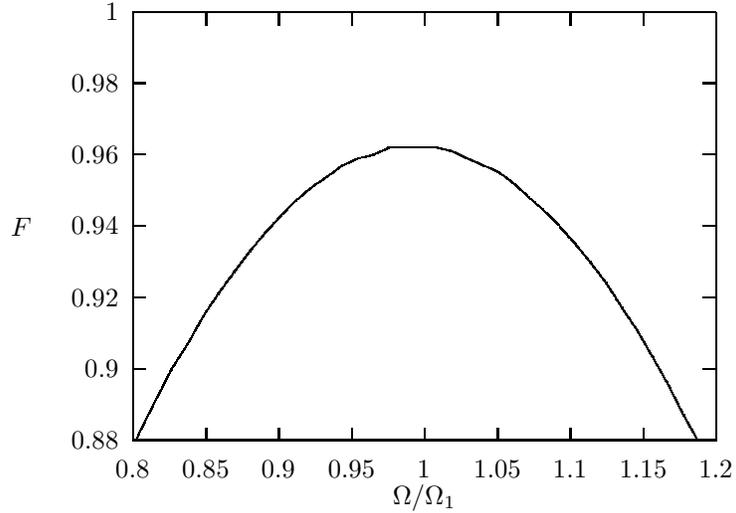
\begin{figure}
\begin{center}
\setlength{\unitlength}{0.240900pt}
\ifx\plotpoint\undefined\newsavebox{\plotpoint}\fi
\sbox{\plotpoint}{\rule[-0.200pt]{0.400pt}{0.400pt}}%
\begin{picture}(1200,809)(0,0)
\font\gnuplot=cmr10 at 10pt
\gnuplot
\sbox{\plotpoint}{\rule[-0.200pt]{0.400pt}{0.400pt}}%
\put(220.0,113.0){\rule[-0.200pt]{4.818pt}{0.400pt}}
\put(198,113){\makebox(0,0)[r]{0.88}}
\put(1116.0,113.0){\rule[-0.200pt]{4.818pt}{0.400pt}}
\put(220.0,225.0){\rule[-0.200pt]{4.818pt}{0.400pt}}
\put(198,225){\makebox(0,0)[r]{0.9}}
\put(1116.0,225.0){\rule[-0.200pt]{4.818pt}{0.400pt}}
\put(220.0,337.0){\rule[-0.200pt]{4.818pt}{0.400pt}}
\put(198,337){\makebox(0,0)[r]{0.92}}
\put(1116.0,337.0){\rule[-0.200pt]{4.818pt}{0.400pt}}
\put(220.0,450.0){\rule[-0.200pt]{4.818pt}{0.400pt}}
\put(198,450){\makebox(0,0)[r]{0.94}}
\put(1116.0,450.0){\rule[-0.200pt]{4.818pt}{0.400pt}}
\put(220.0,562.0){\rule[-0.200pt]{4.818pt}{0.400pt}}
\put(198,562){\makebox(0,0)[r]{0.96}}
\put(1116.0,562.0){\rule[-0.200pt]{4.818pt}{0.400pt}}
\put(220.0,674.0){\rule[-0.200pt]{4.818pt}{0.400pt}}
\put(198,674){\makebox(0,0)[r]{0.98}}
\put(1116.0,674.0){\rule[-0.200pt]{4.818pt}{0.400pt}}
\put(220.0,786.0){\rule[-0.200pt]{4.818pt}{0.400pt}}
\put(198,786){\makebox(0,0)[r]{1}}
\put(1116.0,786.0){\rule[-0.200pt]{4.818pt}{0.400pt}}
\put(220.0,113.0){\rule[-0.200pt]{0.400pt}{4.818pt}}
\put(220,68){\makebox(0,0){0.8}}
\put(220.0,766.0){\rule[-0.200pt]{0.400pt}{4.818pt}}
\put(335.0,113.0){\rule[-0.200pt]{0.400pt}{4.818pt}}
\put(335,68){\makebox(0,0){0.85}}
\put(335.0,766.0){\rule[-0.200pt]{0.400pt}{4.818pt}}
\put(449.0,113.0){\rule[-0.200pt]{0.400pt}{4.818pt}}
\put(449,68){\makebox(0,0){0.9}}
\put(449.0,766.0){\rule[-0.200pt]{0.400pt}{4.818pt}}
\put(564.0,113.0){\rule[-0.200pt]{0.400pt}{4.818pt}}
\put(564,68){\makebox(0,0){0.95}}
\put(564.0,766.0){\rule[-0.200pt]{0.400pt}{4.818pt}}
\put(678.0,113.0){\rule[-0.200pt]{0.400pt}{4.818pt}}
\put(678,68){\makebox(0,0){1}}
\put(678.0,766.0){\rule[-0.200pt]{0.400pt}{4.818pt}}
\put(793.0,113.0){\rule[-0.200pt]{0.400pt}{4.818pt}}
\put(793,68){\makebox(0,0){1.05}}
\put(793.0,766.0){\rule[-0.200pt]{0.400pt}{4.818pt}}
\put(907.0,113.0){\rule[-0.200pt]{0.400pt}{4.818pt}}
\put(907,68){\makebox(0,0){1.1}}
\put(907.0,766.0){\rule[-0.200pt]{0.400pt}{4.818pt}}
\put(1022.0,113.0){\rule[-0.200pt]{0.400pt}{4.818pt}}
\put(1022,68){\makebox(0,0){1.15}}
\put(1022.0,766.0){\rule[-0.200pt]{0.400pt}{4.818pt}}
\put(1136.0,113.0){\rule[-0.200pt]{0.400pt}{4.818pt}}
\put(1136,68){\makebox(0,0){1.2}}
\put(1136.0,766.0){\rule[-0.200pt]{0.400pt}{4.818pt}}
\put(220.0,113.0){\rule[-0.200pt]{220.664pt}{0.400pt}}
\put(1136.0,113.0){\rule[-0.200pt]{0.400pt}{162.126pt}}
\put(220.0,786.0){\rule[-0.200pt]{220.664pt}{0.400pt}}
\put(45,449){\makebox(0,0){$F$}}
\put(678,23){\makebox(0,0){$\Omega/\Omega_1$}}
\put(220.0,113.0){\rule[-0.200pt]{0.400pt}{162.126pt}}
\multiput(226.59,113.00)(0.488,1.088){13}{\rule{0.117pt}{0.950pt}}
\multiput(225.17,113.00)(8.000,15.028){2}{\rule{0.400pt}{0.475pt}}
\multiput(234.58,130.00)(0.497,1.006){47}{\rule{0.120pt}{0.900pt}}
\multiput(233.17,130.00)(25.000,48.132){2}{\rule{0.400pt}{0.450pt}}
\multiput(259.58,180.00)(0.496,0.984){43}{\rule{0.120pt}{0.883pt}}
\multiput(258.17,180.00)(23.000,43.168){2}{\rule{0.400pt}{0.441pt}}
\multiput(282.58,225.00)(0.497,0.782){47}{\rule{0.120pt}{0.724pt}}
\multiput(281.17,225.00)(25.000,37.497){2}{\rule{0.400pt}{0.362pt}}
\multiput(307.58,264.00)(0.497,0.904){47}{\rule{0.120pt}{0.820pt}}
\multiput(306.17,264.00)(25.000,43.298){2}{\rule{0.400pt}{0.410pt}}
\multiput(332.58,309.00)(0.496,0.741){43}{\rule{0.120pt}{0.691pt}}
\multiput(331.17,309.00)(23.000,32.565){2}{\rule{0.400pt}{0.346pt}}
\multiput(355.58,343.00)(0.497,0.681){47}{\rule{0.120pt}{0.644pt}}
\multiput(354.17,343.00)(25.000,32.663){2}{\rule{0.400pt}{0.322pt}}
\multiput(380.58,377.00)(0.497,0.661){47}{\rule{0.120pt}{0.628pt}}
\multiput(379.17,377.00)(25.000,31.697){2}{\rule{0.400pt}{0.314pt}}
\multiput(405.58,410.00)(0.496,0.609){43}{\rule{0.120pt}{0.587pt}}
\multiput(404.17,410.00)(23.000,26.782){2}{\rule{0.400pt}{0.293pt}}
\multiput(428.58,438.00)(0.497,0.537){49}{\rule{0.120pt}{0.531pt}}
\multiput(427.17,438.00)(26.000,26.898){2}{\rule{0.400pt}{0.265pt}}
\multiput(454.58,466.00)(0.496,0.521){41}{\rule{0.120pt}{0.518pt}}
\multiput(453.17,466.00)(22.000,21.924){2}{\rule{0.400pt}{0.259pt}}
\multiput(476.00,489.58)(0.591,0.496){41}{\rule{0.573pt}{0.120pt}}
\multiput(476.00,488.17)(24.811,22.000){2}{\rule{0.286pt}{0.400pt}}
\multiput(502.00,511.58)(0.738,0.495){31}{\rule{0.688pt}{0.119pt}}
\multiput(502.00,510.17)(23.572,17.000){2}{\rule{0.344pt}{0.400pt}}
\multiput(527.00,528.58)(0.678,0.495){31}{\rule{0.641pt}{0.119pt}}
\multiput(527.00,527.17)(21.669,17.000){2}{\rule{0.321pt}{0.400pt}}
\multiput(550.00,545.58)(1.156,0.492){19}{\rule{1.009pt}{0.118pt}}
\multiput(550.00,544.17)(22.906,11.000){2}{\rule{0.505pt}{0.400pt}}
\multiput(575.00,556.59)(2.208,0.482){9}{\rule{1.767pt}{0.116pt}}
\multiput(575.00,555.17)(21.333,6.000){2}{\rule{0.883pt}{0.400pt}}
\multiput(600.00,562.58)(1.062,0.492){19}{\rule{0.936pt}{0.118pt}}
\multiput(600.00,561.17)(21.057,11.000){2}{\rule{0.468pt}{0.400pt}}
\multiput(696.00,571.93)(2.299,-0.482){9}{\rule{1.833pt}{0.116pt}}
\multiput(696.00,572.17)(22.195,-6.000){2}{\rule{0.917pt}{0.400pt}}
\multiput(722.00,565.92)(1.015,-0.492){19}{\rule{0.900pt}{0.118pt}}
\multiput(722.00,566.17)(20.132,-11.000){2}{\rule{0.450pt}{0.400pt}}
\multiput(744.00,554.92)(1.203,-0.492){19}{\rule{1.045pt}{0.118pt}}
\multiput(744.00,555.17)(23.830,-11.000){2}{\rule{0.523pt}{0.400pt}}
\multiput(770.00,543.92)(1.156,-0.492){19}{\rule{1.009pt}{0.118pt}}
\multiput(770.00,544.17)(22.906,-11.000){2}{\rule{0.505pt}{0.400pt}}
\multiput(795.00,532.92)(0.678,-0.495){31}{\rule{0.641pt}{0.119pt}}
\multiput(795.00,533.17)(21.669,-17.000){2}{\rule{0.321pt}{0.400pt}}
\multiput(818.00,515.92)(0.542,-0.496){43}{\rule{0.535pt}{0.120pt}}
\multiput(818.00,516.17)(23.890,-23.000){2}{\rule{0.267pt}{0.400pt}}
\multiput(843.00,492.92)(0.567,-0.496){41}{\rule{0.555pt}{0.120pt}}
\multiput(843.00,493.17)(23.849,-22.000){2}{\rule{0.277pt}{0.400pt}}
\multiput(868.00,470.92)(0.498,-0.496){43}{\rule{0.500pt}{0.120pt}}
\multiput(868.00,471.17)(21.962,-23.000){2}{\rule{0.250pt}{0.400pt}}
\multiput(891.58,446.73)(0.497,-0.559){47}{\rule{0.120pt}{0.548pt}}
\multiput(890.17,447.86)(25.000,-26.863){2}{\rule{0.400pt}{0.274pt}}
\multiput(916.58,418.56)(0.496,-0.609){43}{\rule{0.120pt}{0.587pt}}
\multiput(915.17,419.78)(23.000,-26.782){2}{\rule{0.400pt}{0.293pt}}
\multiput(939.58,390.39)(0.497,-0.661){47}{\rule{0.120pt}{0.628pt}}
\multiput(938.17,391.70)(25.000,-31.697){2}{\rule{0.400pt}{0.314pt}}
\multiput(964.58,356.99)(0.497,-0.782){47}{\rule{0.120pt}{0.724pt}}
\multiput(963.17,358.50)(25.000,-37.497){2}{\rule{0.400pt}{0.362pt}}
\multiput(989.58,318.13)(0.496,-0.741){43}{\rule{0.120pt}{0.691pt}}
\multiput(988.17,319.57)(23.000,-32.565){2}{\rule{0.400pt}{0.346pt}}
\multiput(1012.58,283.71)(0.497,-0.869){49}{\rule{0.120pt}{0.792pt}}
\multiput(1011.17,285.36)(26.000,-43.356){2}{\rule{0.400pt}{0.396pt}}
\multiput(1038.58,238.64)(0.496,-0.891){41}{\rule{0.120pt}{0.809pt}}
\multiput(1037.17,240.32)(22.000,-37.321){2}{\rule{0.400pt}{0.405pt}}
\multiput(1060.58,199.33)(0.497,-0.986){49}{\rule{0.120pt}{0.885pt}}
\multiput(1059.17,201.16)(26.000,-49.164){2}{\rule{0.400pt}{0.442pt}}
\multiput(1086.58,148.64)(0.496,-0.891){41}{\rule{0.120pt}{0.809pt}}
\multiput(1085.17,150.32)(22.000,-37.321){2}{\rule{0.400pt}{0.405pt}}
\put(623.0,573.0){\rule[-0.200pt]{17.586pt}{0.400pt}}
\end{picture}
\end{center}
\caption{The fidelity $F$ of the state prepared by a laser pulse with Rabi frequency $\Omega$ and length $\pi / ( \sqrt{2} \Omega_1 )$ with $\Omega_1=0.25 \, {\rm Im}\, C$. Also, if the length of the laser pulse differs from its ideal value $\pi / ( \sqrt{2} \Omega )$ we find that the fidelity is still above $88 \%$.}
\end{figure}

One the other hand, one could expect the state preparation to be more sensitive to variations in the laser detuning, given that the level shift strongly 
depends from the distance $r$ of the ions, as shown in Figure 2. But we see that if the detuning of the laser differs from $\hbar {\rm Im} \,C/2$, the state $|s \rangle$ is still prepared to a good approximation, as illustrated in Figure 7.

\begin{figure}
\begin{center}
\input{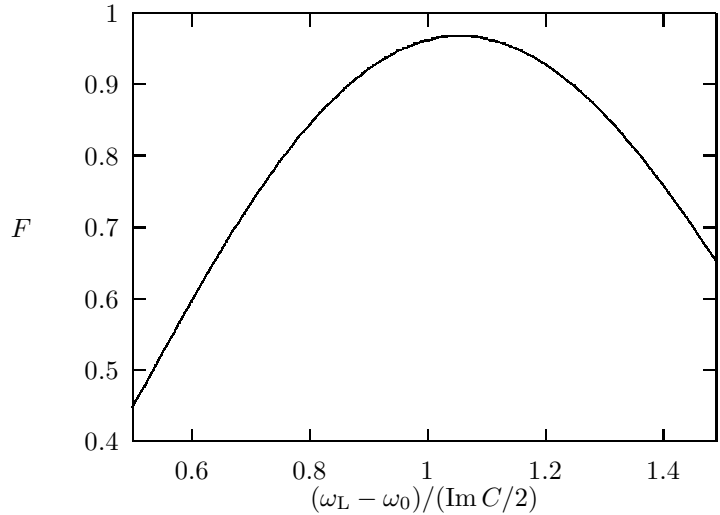}
\end{center}
\caption{Population of level $\ket{s}$ after a laser pulse of duration $T_\pi$ $(\Omega_1 =0.25 \, {\rm Im}\,C)$, for different values of the laser detuning 
$\omega_{\rm L}-\omega_0$. The fidelity of the prepared state $F$ is maximal, if the laser frequency is somewhat above the resonance frequency of the $g$-$s$ transition and is still high if the detuning varies from its ideal value by up to $20\%$.}
\end{figure}

The fidelity of the prepared state cannot become equal to 1, because there is always also a slightly (even though detuned) pumping to state $|e \rangle$. Therefore it is not surprising that the fidelity of the prepared state is maximal for laser detuning slightly above ${\rm Im}\,C/2$. In this case one has less excitation of level $e$ and this advantage is not totally compensated by the worse coupling of the laser on the $g$-$s$ transition.

\subsection{Generation of the antisymmetric state $\ket{a}$.}

In a similar way, if one sets $\Omega_1=-\Omega_2$ and tunes the laser frequency to $\omega_0-{\rm Im} \, C/2$, the maximally entangled state $|a \rangle$ is generated by the laser pulse, since the Hamiltonian $H_{\rm I}$ becomes
\begin{eqnarray} \label{39}
H_{\rm I} &=& 
- \frac{\hbar \, \Omega_1}{\sqrt{2}} \, \Big( |g \rangle \langle a| 
- {\rm e}^{-{\rm i} {\rm Im}\,C t} \, |a \rangle \langle e| \Big) 
+ {\rm h.c.} 
\end{eqnarray}
In this case, only the transition to level $a$ is excited, to a good approximation, as long as the Rabi frequency $\Omega_1$ is not too large and ${\rm e}^{-{\rm i} {\rm Im}\,C t}$ can be considered as a fast oscillating term.

The only problem which arises now is that a phase difference of $\pi$ between the Rabi frequencies $\Omega_1$ and $\Omega_2$ cannot be achieved by applying only a single laser field, because the distance between the ions is much smaller than the wavelength $\lambda_0$. One possibility is to use a standing wave configuration, where both ions are placed symmetrically around a node of the laser field \cite{general}. Then the Rabi frequency at ion 1 becomes, analogous to Eq. (\ref{32}), but summing now over two laser fields,
\begin{eqnarray} \label{310}
\Omega_1 &=& \frac{e}{\hbar} \, {\bf D}_{01} \cdot {\bf E}_{0} 
\Big( {\rm e}^{{\rm i} k_{\rm L} r/2}-{\rm e}^{-{\rm i} k_{\rm L} r/2} \Big)
~=~ \frac{e}{\hbar} \, {\bf D}_{01} \cdot {\bf E}_{0} 
\, 2 {\rm i} \sin k_{\rm L} r/2,
\end{eqnarray}
where ${\bf E}_{0}$ is the amplitude of the laser coming in from the side of ion 1 at the centre between the two ions. For $k_0r \approx  k_{\rm L}r \approx 0.2$ this Rabi frequency is still about one fifth of the Rabi frequency $\Omega_1$ of the running wave used in the previous subsection to prepare the entangled state $|s \rangle$. 

\section{Realisation of a CNOT quantum gate.}

A CNOT is an operation where the value of the second qubit is changed depending on the value of the first qubit. Using the basis ${\cal B}$ of Eq. (\ref{11}), this operation is equal to an unitary operation which does not affect the states $|g \rangle$ and $(|s \rangle + |a \rangle)/\sqrt{2}$, while one has
\begin{eqnarray} \label{42}
(|s\rangle-|a \rangle)/\sqrt{2} & \longrightarrow & |e \rangle, \nonumber \\
|e \rangle & \longrightarrow & 
(|s \rangle - |a \rangle)/\sqrt{2}.
\end{eqnarray}
In the following we will discuss how this operation can be realised to a very good approximation in the present scheme. Let us first assume that the interaction Hamiltonian of the system is given by 
\begin{eqnarray} \label{43}
H_{\rm I} &=& \frac{\hbar \Omega}{2} \, 
\Big( |e \rangle (\langle s| + \langle a|)/\sqrt{2}  + {\rm h.c.} \Big).
\end{eqnarray}
Then one has
\begin{eqnarray} \label{46}
U_{\rm I}(t,0) \, (|s \rangle - |a \rangle)/\sqrt{2} 
&=& \cos (\Omega t/2) \, (|s \rangle - |a \rangle)/\sqrt{2} 
-{\rm i} \, \sin (\Omega t/2) \, |e \rangle , \nonumber \\
U_{\rm I}(t,0) \, |e \rangle
&=& -{\rm i} \, \sin (\Omega t/2) \,
(|s \rangle - |a \rangle)/\sqrt{2} + \cos (\Omega t/2) \, |e \rangle
\end{eqnarray}
and the population of the states $|e \rangle$ and $(|s \rangle - |a \rangle)/\sqrt{2}$ are exchanged after a pulse of the length $T_\pi=\pi/\Omega$. On the other hand the states $|g \rangle$ and $(|s \rangle + |a \rangle)/\sqrt{2}$ are eigenstates of the Hamiltonian with a zero eigenvalue and are not affected. Therefore a CNOT operation can be performed by the given interaction. 

\begin{minipage}{6.54truein}
\begin{figure}[htb]
\begin{center}
\epsfxsize11.0cm
\centerline{\epsfbox{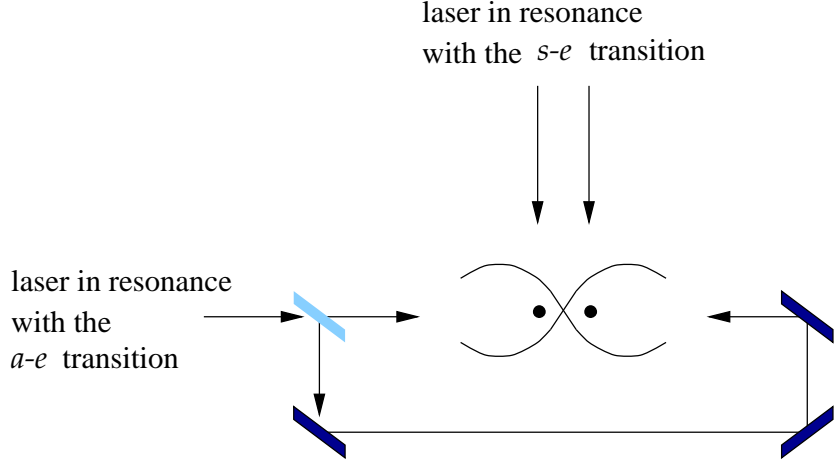}}
\end{center}
\caption{\small Experimental set-up for the realisation of a CNOT operation. 
The two ions are placed symmetrically around the node of a standing laser wave which is in resonance with the $a$-$e$ transition. Another laser with a wave vector perpendicular to the line connecting both ions is employed to excite the $s$-$e$ transition. The Rabi frequencies have to be chosen such that they differ only in sign.} 
\end{figure}
\end{minipage}

\vspace*{0.5cm}
The Hamiltonian in Eq. (\ref{43}) can be realised to a good approximation by the configuration shown in Figure 8, where the laser fields couple to the $s$-$e$ and the $a$-$e$ transition only. The first transition is driven by a laser with the detuning $\omega_{\rm L}-\omega_0=-{\rm Im}\,C/2$ and a wave vector ${\bf k}_{\rm L}$ perpendicular to the line connecting both ions. Its Rabi frequency at ion 1 is analogous to Eq.~(\ref{o1}) given by
\begin{eqnarray} \label{47}
\Omega_{1\rm r} &=& \frac{e}{\hbar} \, {\bf D}_{01} \cdot 
{\bf E}_{0\rm r}.
\end{eqnarray}
The $a$-$e$ transition can be excited by a standing laser wave with the frequency  $\omega_0+{\rm Im} \, C/2$. Analogous to Eq. (\ref{310}) its Rabi frequency $\Omega_{1\rm s}$ is 
\begin{eqnarray} \label{48}
\Omega_{1\rm s} &=& 
\frac{e}{\hbar} \, {\bf D}_{01} \cdot {\bf E}_{0\rm s} 
\, 2 {\rm i} \sin k_{\rm L} r/2.
\end{eqnarray}
${\bf E}_{0\rm r}$ and ${\bf E}_{0\rm s}$, respectively, denote the amplitude of the corresponding laser (in the second case coming in from the side of ion 1) at the centre between the two ions. Then the Hamiltonian of the system with respect to $H_0$ of Eq.~(\ref{29}) is according to Eq.~(\ref{34}) given by
\begin{eqnarray} \label{49}
H_{\rm I} &=&  \frac{\hbar}{\sqrt{2}} \, \Big(
\Omega_{1\rm r} \, |s \rangle \langle e| 
+ \Omega_{1\rm s} \, |a \rangle \langle e| 
+ {\rm e}^{-{\rm i} {\rm Im}\,Ct} |g \rangle \langle s|
- {\rm e}^{{\rm i} {\rm Im}\,Ct} |g \rangle \langle a| \Big) + {\rm h.c.} 
\end{eqnarray}
This Hamiltonian differs from the one in Eq.~(\ref{43}) only by some oscillating terms if the condition 
\begin{eqnarray} \label{410}
\Omega_{1\rm s} &=& - \Omega_{1\rm r}
\end{eqnarray}
is fulfilled. This can be achieved by varying the phase between the amplitudes ${\bf E}_{0\rm s}$ and ${\bf E}_{0\rm r}$ and one gets Eq.~(\ref{43}) with
\begin{eqnarray} \label{411}
\Omega & \equiv& 2 \Omega_{1\rm r}.
\end{eqnarray}
Figure 9 shows how the population of various states varies in time, if the system is initially prepared in the excited state $|e \rangle$. The results have been obtained numerically taking the full Hamiltonian of Eq.~(\ref{49}) into account. As seen in the previous section the oscillating terms do not contribute if the detuning of the lasers is strong enough, e.g.~for instance one can take $\Omega_{1\rm s}={\rm Im} \, C/4$ as in Figure 9. Then duration of a CNOT operation is given by 
\begin{equation}
T_\pi = \pi/(2\Omega_{1 \rm r}) = 0.017 /A~. 
\end{equation}

\begin{figure}
\begin{center}
\input{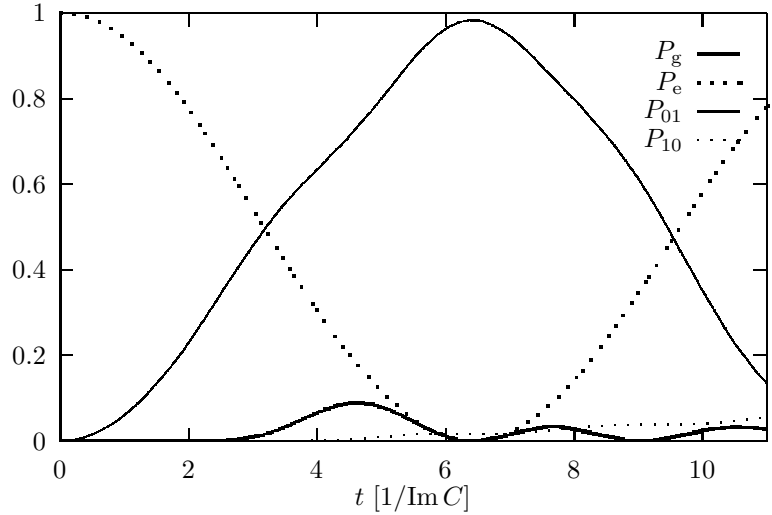}
\end{center}
\caption{The time dependence of the population of the states $|g \rangle$, $|e \rangle$, $|01 \rangle = (|s \rangle -|a \rangle)/\sqrt{2}$ and $|10 \rangle = (|s \rangle +|a \rangle)/\sqrt{2}$, if the ions are initially prepared in $|e \rangle$. The Rabi frequency of the running laser $\Omega_{1\rm r}$ is equal to $0.25\, {\rm Im}\,C$, while at the same time a standing wave with the Rabi frequency $-\Omega_{1\rm r}$ is applied to drive the $a$-$e$ transition.
After the time $T_\pi=\pi/(2\Omega_{1\rm r})$ the ions are to a good approximation in the state expected for a CNOT operation.}
\end{figure}

\section{Realisation of a single ion gate operation.}

One would expect that for a single ion operation it is necessary to focus a laser on one of the ions. However, because the ions are well localised within a wavelength, this is here experimentally impossible. Therefore we propose another way to realise single ion operations. Considering the interaction of only one ion with a laser field, its basis vectors $|0 \rangle$ and $|1 \rangle$ are changed in the following way 
\begin{eqnarray} \label{51}
|0 \rangle & \longrightarrow & \cos (\Omega t/2) \, |0 \rangle  
-{\rm i} \sin (\Omega t/2) \, |1 \rangle,  \nonumber \\ 
|1 \rangle & \longrightarrow & -{\rm i} \sin (\Omega t/2) \, |0 \rangle
+ \cos (\Omega t/2) \, |1 \rangle.
\end{eqnarray}
The state of the other ion is not changed by the transformation. 
Using the basis ${\cal B}$ of Eq. (\ref{11}) this operation corresponds for instance to the following two-step process. The first step is equal to an unitary transformation, where the states $|e \rangle$ and $(|s \rangle + |a \rangle)/\sqrt{2}$ are unchanged, while 
\begin{eqnarray} \label{52}
|g \rangle & \longrightarrow & \cos (\Omega t/2) \, |g \rangle  
-{\rm i} \sin (\Omega t/2) \, (|s \rangle - |a \rangle)/\sqrt{2}, \nonumber \\ 
(|s \rangle - |a \rangle)/\sqrt{2} & \longrightarrow & -{\rm i} 
\sin (\Omega t/2) \, |g \rangle + \cos (\Omega t/2) \, 
(|s \rangle - |a \rangle)/\sqrt{2} .
\end{eqnarray}
In the second step one has to ensure that 
\begin{eqnarray} \label{53}
|e \rangle & \longrightarrow & \cos (\Omega t/2) \, |e \rangle  
-{\rm i} \sin (\Omega t/2) \, (|s \rangle + |a \rangle)/\sqrt{2}, \nonumber \\ 
(|s \rangle + |a \rangle)/\sqrt{2} & \longrightarrow & 
-{\rm i} \sin (\Omega t/2) \, |e \rangle
+ \cos (\Omega t/2) \, (|s \rangle + |a \rangle)/\sqrt{2} ,
\end{eqnarray}
while $|g \rangle$ and $(|s \rangle - |a \rangle)/\sqrt{2}$ are not affected. These steps generate the CNOT operation described in Section IV and can be realised in a similar way.
Laser fields have to be applied which couple only to the $g$-$s$ and the $g$-$a$ transitions. As in Section IV the corresponding Rabi frequencies can be chosen in such a way that the state $(|s \rangle + |a \rangle)/\sqrt{2}$ is not affected. The necessary condition is 
\begin{eqnarray} \label{54}
\Omega_{1\rm r} &=& -\Omega_{1\rm s} ~\equiv~ \Omega/2,
\end{eqnarray}
where $\Omega_{1\rm r}$ and $\Omega_{1\rm s}$ are the Rabi frequencies of the laser field in resonance with the $g$-$s$ and the $g$-$a$ transition, respectively. The population of $|e \rangle$ is not changed during this process, because no laser couples to this state. In the second step, to perform transformation (\ref{53}), only the $s$-$e$ and the $a$-$e$ transition are driven. This can be achieved by choosing appropriate detunings of the applied fields. The corresponding Rabi frequencies have to fulfill the condition
\begin{eqnarray} \label{55}
\Omega_{1 \rm r} &=& \Omega_{1 \rm s} ~\equiv~ \Omega/2.
\end{eqnarray}
Then the state $(|s \rangle - |a \rangle)/\sqrt{2}$ is an eigenstate with zero eigenvalue of the interaction Hamiltonian.

In an analogous way single ion operation on ion 2 can be performed. Again, to realise this operation is more complicated than in the case of well separated ions, but single ion operations are possible in the system considered here, as the level shifts of states $|s \rangle$ and $|a \rangle$ allow us to excite all possible transitions separately.

\section{Conclusions.} 

In this paper the idea of quantum gate operation with two dipole-interacting ions in a linear ion trap has been examined in detail. We showed that it is possible to create a maximally entangled state simply by applying a single laser pulse in a standing or running wave configuration. In addition we discussed how to implement conditional dynamics (a CNOT gate) and single qubit operations. 

To do this the level shift induced by the dipole-dipole interaction between the ions in the trap has been used, which allows us to address every single transition in the system separately. The level shift has to be much larger than the decay rate of the levels and the applied Rabi frequencies.
Therefore the two-level ions have to be very close, e.g.~much closer than the wavelength of their emitted light. This requires ions with very long wavelengths (Rydberg ions), or one has to use ion traps with very strong confining trap potentials. The strength and weakness from the point of view of practical realization have been discussed. \\[1cm]

{\bf Acknowledgements}
\noindent
We thank J.~Hoeffges, J.~Steinbach and W.~Lange for useful comments on the subject of this paper. This work was supported by a Feodor Lynen grant of the Alexander 
von Humboldt Foundation, the European Community TMR Networks ERB 406PL95-1412 and ERBFMRXCT96066, the European Science Foundation, the UK Engineering 
and Physical Sciences Research Council und the Leverhulme Trust.

\end{document}